# Building MultiView Analyst Profile From Multidimensional Query Logs: From Consensual to Conflicting Preferences

**Eya Ben Ahmed[1], Ahlem Nabli[2] and Faïez Gargouri[3]**

**[1] Computer Science Department, Higher Institute of Management of Tunis,
Tunis, Tunisia**

**[2] Computer Science Department, Faculty of Sciences of Sfax,
Sfax, Tunisia**

**[3] Computer Science Department, Higher Institute of Computer Science and Multimedia of Sfax,
Sfax, Tunisia**

## Abstract

In order to provide suitable results to the analyst needs, user preferences summarization is widely used in several domains. In this paper, we introduce a new approach for user profile construction from OLAP query logs. The key idea is to learn the user's preferences by drawing the evidence from OLAP logs. In fact, the analyst preferences are clustered into three main pools : (i) consensual or non conflicting preferences referring to same preferences for all analysts; (ii) semi-conflicting preferences corresponding to similar preferences for some analysts; (iii) conflicting preferences related to disjoint preferences for all analysts. To build generic and global model accurately describing the analyst, we enrich the obtained characteristics through including several views, namely the personal view, the professional view and the behavioral view. After that, the multiview profile extracted from multidimensional database can be annotated.

## 1. Introduction

Data warehouses store a large amount of information which are analyzed in order to support strategic decision makers. OLAP analyses consist in exploring interactively the data warehouse using navigational operations. To better fit the analyst's needs, several complicated operations may be performed. Generally, some analyses are usually made by the same decision makers. Despite the diversity of the analysts' intentions, existing OLAP technology provides regularly the same results for the same keyword queries. The main reason behind this is that the search process is made out of the user features.

In fact, collecting relevant user interests and main user preferences in a user profile, may efficiently enhance support personalization and user-centric adaptivity. The notion of user profiling has been introduced in order to personalize applications so as to be tailored to the user needs. The user profile may contain different types of information: *personal data* such as identity, demographic data; *professional data* such as position/function, principal responsibilities, role and duties; and finally *behavioral data* mainly related to the data warehouse schema preferences.

Accordingly, our work focuses on the multiview analyst profile building from OLAP logs: First, we prepare the text in a preprocessing stage. Second, we cluster behavioral information in consensual, semi-conflicting and conflicting preferences. Then, we generate a generic user profile through its enrichment by mandatory views. Finally, the derived user profile may be annotated.

The rest of the paper is organized as follows: Section (2) introduces the work related to user profile modeling. In section (3), we present our approach for user profile construction and annotation. In order to validate our contribution, we describe the three steps we followed to carry out the OLAPAnalystProfile system. Experimental results evaluating the efficiency of our system are reported in section (4). In section (5) we conclude our work and briefly outline future work.

## 2. Related Work

In this section, we focus on the various research work closely related to the domain of the user profile content and the user profile modeling in data warehouse area.

Table 1: Comparison of user profile modeling approaches

| Method | IR | DW | Explicit | Implicit | Ontological | Non ontological | Atemporal | Short-term | Long-term | Conflicting | Non conflicting |
|---|---|---|---|---|---|---|---|---|---|---|---|
| | | | Acquisition | | Semantic | | Term | | | Conflict | |
| Gowan [10] | X | | X | | | X | X | | | | X |
| Sieg et al. [16, 17] | X | | | X | X | | | X | X | | X |
| Liu et al. [9] | X | | | X | X | | X | | | | X |
| Challam et al. [3] | X | | | X | X | | X | | | | X |
| Cherniack et al. [4] | X | | | X | | X | X | | | | X |
| Bouze-ghoub, Kostadinov [2] | X | | X | | | X | | X | X | | X |
| Ravat et al. [11, 12] | | X | X | | | X | | | X | | X |
| Jerbi et al. [7, 8] | | X | X | | | X | | X | X | | X |
| Rizzi and Golfarelli [6, 13, 14] | | X | X | | | X | | | X | | X |
| **Our proposal** | | X | | X | X | | X | | | X | |

## 2.1 User profile content

Several definitions of user profile are proposed. According to its representation in Information Retrieval (IR) area, we distinguish the following definitions of the user profile content:

- As *weighted keyword vectors* [10]. Generally, such profile is represented using vectorial representation. For example, the user profile is composed of two keywords weighted (*status*, *query*) as follows (*i.e.* {*status*, 0.7; *query*, 0.8});
- As *semantically weighted ontological concepts* [16] combining the user's interests and Yahoo concept hierarchy.
  According to [3], the user profile is a set of weighted concepts selected from the ODP ontology. According to Liu et al. [9], the user profile consists in a set of categories and for each category, a set of terms (keywords) with weights is defined. The weight of a term in a category reflects the significance of the term on representing the user's interest in that category.

- Using *the utility notion*. In fact, the profile specification is broken into two parts [4]: the Domain clause (DOMAIN) defines and names sets of objects of interest (domain sets); and the utility clause (UTILITY) specifies the relative values of objects contained in each domain set.
- Using *several dimensions* [2], such as the user interest, the context of the launched query, the accurate level of quality, the interactions history and the different preferences on these dimensions.

## 2.2 User profile modeling in data warehouses

Taking aggregation into account, the user profile content is restricted to expressed preferences on schema rather than on instances as commonly done in Data Warehouses (DW) [1].

Ravat et al. [11,12] proposed a conceptual model of user profile based on multidimensional concepts (fact, dimension, hierarchy, measure, parameter or attribute). To assign priority weights to attributes of a multidimensional schema, the personalization rules are described using the Condition-Action formalism.

Accordingly, an OLAP query language adapted to the personalization context is proposed. The weights are taken into account during OLAP analyses. In addition, the proposed algebra contains OLAP operators allowing the drilling, rotations, selection, ordering, aggregation and modification operations.

Jerbi et al [7,8] propose a context-aware OLAP Preference model which is defined on MDB schema. Using a qualitative approach, the OLAP preferences are modeled and closely depend on user analysis context (c.f. figure 1). That's why a conceptual model of OLAP context is conceived using an arborescence of OLAP analysis elements.

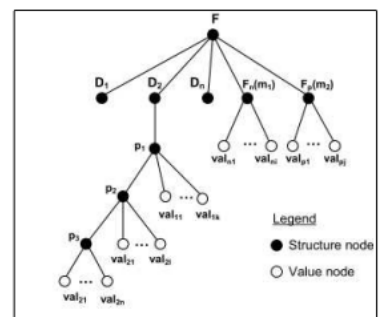

Fig. 1. OLAP analysis context Tree.

Rizzi [13] introduce MyOLAP approach where preferences are expressed using a strict partial order and are made through the first-order logic formulas.

For instance, an illustrative example of formulated preference is presented in the following. Let's consider A a set of attributes belonging to the domain *dom*(A). A preference P is a strict partial order $P = (A, < P) \in dom(A)$; $x <P y$ is interpreted as *I like y better than x.*

Formulated on schema, the preferences concern not only dimensional attributes but also measures, and group-by sets. A preference algebra has been introduced to manage the different relation between preferences using specific defined operators.

The process of user profiling in data warehouse area is strictly restricted to schema personalization. Indeed, no added value information that can fundamentally orient the user preferences are taken into consideration, namely professional information such as current function, abilities and disabilities, etc. As shown by table 1, the main distinctive feature of our work is the automatic creation of analyst profile from his history in data warehouses including several views and handling the conflict aspect.

## 3. The Proposed Approach of Analyst Profile Construction And Annotation

The automatic process of OLAPAnalystProfile system is carried out in three stages (cf. Figure 2):

- **Preprocessing**: of OLAP log queries, it consists, on the one hand, in the session and text segmentation and, on the other hand, in the identification of entities.
- **Generic profile construction**: it consists in clustering of preferences on three main pools: (i) *consensual preferences*; (ii) *semi-conflicting* preferences and (iii) *conflicting preferences*. We propose in this stage a new similarity measurement between the preferences. After that, such preferences are enriched using mandatory information to build generic and global profile.
- **Profile Annotation**: in order to facilitate the classification, adding information in the profile-content, correlating two preferences or investigation of future actions in the created profile, the annotation may take several forms, such as element of the data warehouse schema and its frequency.

### 3.1 Preprocessing

In the case study, the preprocessing allows the text segmentation into sentences and the delimitation of entities.

**Segmentation**: is the determination of the sessions boundaries, on the one hand, and sentences borders, on the other hand. The existing tools can be categorized as follows: (i) some of them take into account all typographical markers, (ii) other tools are backboned on linguistic bases (i.e. the syntactic structure of a sentence or the significance of each typographical marker).

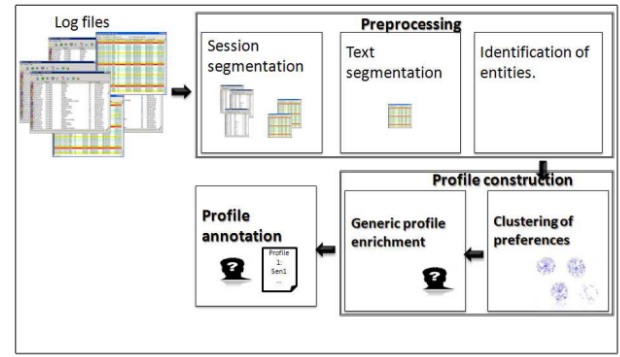

Fig. 2. Architecture of the analyst profile construction and annotation system.

Taking benefit from the structure of an MDX query, we have developed our own segmentor relying on both punctuation marks and the MDX query form. The result of the text segmentation of OLAP log file is shown by Figure 3.

```
<?xml version="1.0" encoding="UTF-8"?>
<Log session = "SalesManager1">
<Query id = "1">
<Columns id = "1"> Sales Amount </Columns>
<Rows id = "1"> Year=2010 </Rows>
<Rows id = "2"> Year=2011 </Rows>
<Cube id ="1"> Sales </Cube>
<Condition id = "1"> [Tunisia].[Tunis] </Condition>
</Query>
<!--...-
</Log session>
</xml>
```

Fig. 3. Extract of the OLAP log segmentation.

**Named Entities Recognition** : Named Entities are types of specific lexemes referring to an entity of the concrete world in given domain, namely social, medical, economic or geographical area and having a particular name [5]. The entities are identified in the log files by a tag which corresponds to the entity type. The types selected are recognized by rules using the multidimensional schema considered as a dictionary of named entities. Figure 4 presents the same text of Figure 3 after the named entities identification. Initially the position of each term is fixed. Then, each entity is recognized by specifying its attributes.

```
<?xml version="1.0" encoding="UTF-8"?>
<Log session = "SalesManager1">
<Query id = "1">
<Measure id = "1"> Sales Amount </Measure>
<Dimension id = "1"> Time
<Member id = "1"> Year=2010 </Member>
<Member id = "2"> Year=2011 </Member>
</Dimension>
<Cube id = "1"> Sales </Cube>
<Condition id = "1"> [Tunisia].[Tunis] </Condition>
<Dimension id = "1"> Place
<Member id = "1"> Tunisia </Member>
<Member id = "2"> Tunis </Member>
</Dimension>
</Query>
<!--...-->
</Log session>
</xml>
```

Fig. 4. Extract of the OLAP log segmentation after
the named entities recognition.

## 3.2 Generic Profile Construction

This step is composed of clustering of OLAP queries highlighting the analyst preferences on the one side and enrichment of created profile on the other side.

**Clustering Of Preferences**: In this stage, we categorize the behavioral information extracted from the preprocessed log files on three main pools: (i) *consensual or non conflicting preferences* referring to same preferences for all analysts; (ii) *semi-conflicting preferences* corresponding to similar preferences for some analysts; (iii) *conflicting preferences* related to disjoint preferences for all analysts.

We apply the complete link hierarchical clustering algorithm to gather the queries. In fact, this algorithm merges in each step the two closest clusters having the biggest similarity distance. The latter is computed using a similarity measurement between queries.

### A. Similarity Measurement Between Queries

A number of similarity measures is used in the hierarchical clustering to discover the closest pair of documents to merge. Among them, the cosine measure [15] is commonly the most used in document clustering particularly when the number of frequent concepts on each document is drastically different. In addition, this measure is based on the document components and is not sensitive to the document length.

We have chosen the Jaccard distance because it significantly suits the large documents. Thus, we extend this measure to the multidimensional context. The multidimensional Jaccard distance relies on the MDX query structure, particularly on similarity between used facts, measures, dimension attributes, as well as slicer specification members. The later is used in the Where clause and restricts the result data. Any dimension that does not appear on an axis in the SELECT clause can be named on the slicer. The similarity measure is the number of common facts, measures, dimensions and slicer specification members in the two queries divided by the total number of facts, measures, dimensions and slicer specification minus the already computed numerator, it is computed according to the following formula. We suppose

$(A) = C_{Fact(q_i,q_j)} + C_{Measure(q_i,q_j)} + C_{DimensionAttribute(q_i,q_j)} + C_{SSMember(q_i,q_j)}.$

$$J(q_i,q_j) = \frac{(A)}{[\sum_{k=i,j} Fact(q_k) + \sum_{k=i,j} Measure(q_k) + \sum_{k=i,j} DimensionAttribute(q_k) + \sum_{k=i,j} SSMember(q_k)] - [(A)]}$$

with $C_{Fact(q_i,q_j)}$: Common facts of $q_i$ and $q_j$,

$C_{Measure(q_i,q_j)}$: Common measures of $q_i$ and $q_j$,

$C_{DimensionAttribute(q_i,q_j)}$: Common dimension attributes of $q_i$ and $q_j$,

$C_{SSMember(q_i,q_j)}$: Common slicer specification members of $q_i$ and $q_j$.

For example, we consider the two following queries $q_1$ and $q_2$.

*$q_1$: SELECT [Measures].[Sales Amount] ON COLUMNS, [Date].[All]ON ROWS*
*FROM Sales*
*WHERE ([Customer].[France].[Lyon]);*
*$q_2$: SELECT [Measures].[Sales Amount] ON COLUMNS, [Date].[2010], [Date].[2011]ON ROWS*
*FROM Sales*
*WHERE ([Customer].[France].[Lyon]);*

$q_1$ and $q_2$ use the same fact, the same measure and the same slicer specification member. However, $q_1$ uses all dimension attributes of the Date dimension which are 5 and $q_2$ accesses to only two dimension attributes which are 2010 and 2011.

We suppose $(A) = C_{Fact(q_1,q_2)} + C_{Measure(q_1,q_2)} + C_{DimensionAttribute(q_1,q_2)} + C_{SSMember(q_1,q_2)}.$

$$J(q_1,q_2) = \frac{(A)}{[\sum_{k=1,2} Fact(q_k) + \sum_{k=1,2} Measure(q_k) + \sum_{k=1,2} DimensionAttribute(q_k) + \sum_{k=1,2} SSMember(q_k)] - [(A)]}$$

$$= \frac{1+1+2+1}{2+2+2+5-2-(1+1+2+1)} = \frac{5}{13-5} = 0.625$$

### B. Hierarchical Clustering Algorithm

The different construction steps of hierarchical clustering undertaken in our OLAPAnalystProfile are described as follows:

- *Initialization*: Count up the frequencies of each query. Let each query be a cluster; if its frequency is greater than 1, consider only a cluster for each group of repetitive queries;
- *Treatment*: Compute similarity matrix;
- *Assignment*: Merge the two closest clusters based on the two following conditions: *(a)* the maximum of similarity measure using multidimensional Jaccard distance; (b) *the maximum of frequent queries*;
- *Updating*: Update similarity matrix;

- *Iteration*: Repeat steps 3 and 4 until only three clusters remain.

A good clustering method will produce high quality clusters with high intra-cluster similarity and low inter-cluster similarity. To measure the conflict between clusters, we integrate the concept of frequency. Finally, we stop running the algorithm when the number of clusters reaches three which are: (i) *consensual or non conflicting preferences*; (ii) *semi-conflicting preferences* corresponding to similar preferences; (iii) *conflicting preferences*.

For instance, let us consider the four following queries:

$q_1$: *SELECT [Measures].[Sales Amount]ON COLUMNS, [Date].[All]ON ROWS*
*FROM Sales*
*WHERE ([Customer].[France].[Lyon])*
$q_2$: *SELECT [Measures].[Sales Amount] ON COLUMNS, [Date].[2010], [Date].[2011] ON ROWS*
*FROM Sales*
*WHERE ([Customer].[France].[Lyon])*
$q_3$: *SELECT [Measures].[Sales Amount]ON COLUMNS, [Product].[Astradol] ON ROWS*
*FROM Sales*
$q_4$: *SELECT [Measures].[Sales Amount] ON COLUMNS, [Date].[All]ON ROWS*
*FROM Sales*
*WHERE ([Customer].[France].[Lyon])*

Table 2: Similarity matrix

| Distance | $C_1$: Freq($C_1$)=2 | $C_2$: Freq($C_2$)=1 | $C_3$: Freq($C_3$)=1 |
|---|---|---|---|
| $C_1$: Freq($C_1$)=2 | 0 | 0.625 | 0.222 |
| $C_2$: Freq($C_2$)=1 | 0.625 | 0 | 0.333 |
| $C_3$: Freq($C_3$)=1 | 0.222 | 0.333 | 0 |

We start by counting the frequencies of the queries: (i) Freq($q_1$)=2; (ii) Freq($q_2$)=1; (iii) Freq($q_3$)=1; (iv) Freq($q_4$)=2. After that, we assign clusters to queries as follows: (i) $q_1 \Rightarrow C_1$; (ii) $q_2 \Rightarrow C_2$; (iii) $q_3 \Rightarrow C_3$; (iv) $q_4 \Rightarrow C_1$. In fact, the first and the forth queries are merged because they are identical. Then, we compute the similarity matrix shown by the table 2. The similar pair of queries is $C_1$ and $C_2$, at distance 0.625 and $C_1$ is the most frequent. These queries are merged into a single cluster called "$C_1/C_2$". Then we compute the distance from this new compound query to all other queries. In complete link clustering, the rule is that the distance from the compound query to another query is equal to the greatest similarity distance from any member of the cluster to the outside query. So the distance from "$C_1/C_2$" to $C_3$ is chosen to be 0.333 which is the distance from $C_3$ to $C_2$, and so on. After merging $C_1$ with $C_2$, we obtain the matrix illustrated by the table 3. The running example is a sample of our data set. However, in real case when we reach the three clusters, we may finally stop merging.

Table 3: Similarity matrix after merging $C_1$ with $C_2$.

| Distance | $C_1/C_2$ | $C_3$ |
|---|---|---|
| $C_1/C_2$ | 0 | 0.333 |
| $C_3$ | 0.333 | 0 |

**Generic Profile Enrichment:** As outlined by the conceptual modeling of analyst profile shown by figure 5, the behavioral component of the derived profile may be enriched by adding:

- *personal information* such as identity and demographic data. They include the user identity specified using his name, his social security number, etc, demographics identified using his age, his gender, his address, his marital status, his number of children, etc, his professional contacts as well as his credit card number. Generally, such kind of information does not need frequent update.
- *professional information* such as position/function (e.g. sales manager), principal responsibilities (e.g. for sales manager; to achieve the company's goals and to develop the people reporting to them), role (e.g. for the sales manager, to focus on sales; to set sales objectives, forecasting, budgeting, organizing and sales force's recruitment) and duties (e.g. for sales manager; to assign sales territories, or geographic regions to selling personnel; to evaluate the performance of the sales workers; to represent his company at trade association conventions and meetings; to promote his products, etc).

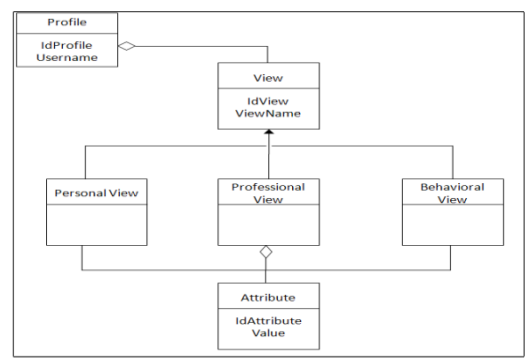

Fig. 5. Conceptual modeling of analyst profile.

## 3.3 Profile Annotation

We continue the enrichment of the profile by other metadata which will be very useful for all ulterior treatment (information retrieval, automatic summarization, storage of the preferences in a database, indexation, etc).

Mainly, we annotate the user profile-content by adding frequency to each clause of behavioral preference. Eventually, we store each preference and each related annotation in a separate database.

For instance, we present an example of profile annotation shown by the figure 6. Indeed, the cluster is annotated through the frequencies of the fact Sales, the measure Sales Amount, the dimension Date and the slicer specification members Customer.France.Lyon and Customer.France.Paris which are respectively 2, 2, 2, 1 and 1.

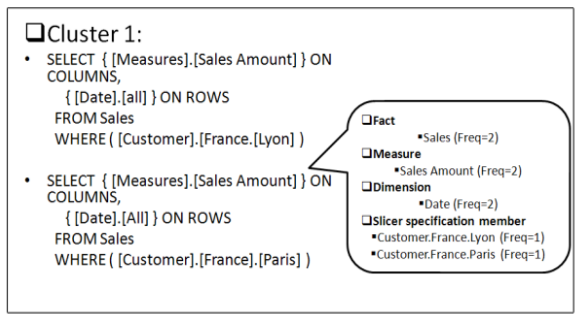

Fig. 6. Generic profile annotation.

## 4. Experimental results

In order to validate our approach, we have implemented our system OLAPAnalystProfile using the java language. In fact, our system contains three modules:

- *Module one*: Preprocessing through sessions segmentations and entities identification in log files;
- *Module two*: Generic profile construction through the clustering of preferences on three pools: (i) *consensual preferences*; (ii) *semi-conflicting preferences*; (iii) *conflicting preferences*; then its enrichment in order to derive an extended profile;
- *Module three*: Annotation of generated profile.

We prepared a corpus of 5000 OLAP queries stored in log file. We used the Weka 3.6.5 edition to apply the decision trees. First, we segment them in sessions then in queries in order to process the log files. Second, we identify the named entities based on the data warehouse schema. Then, we start the construction of profiles through the clustering of queries on three pools: (i) consensual preferences; (ii) semi-conflicting preferences; (iii) conflicting preferences. Hence, a hierarchical clustering algorithm is applied and an innovative extension of Jaccard measure is proposed in the multidimensional context. Then, an enrichment of the generated profile is performed through adding personal and professional information to behavioral ones. Finally, such a profile may be annotated using the frequency of each clause of behavioral preferences.

Our training set is presented by a part of group of queries being the output of the preprocessing step. Such a set is annotated by an expert. For each OLAP query, the default value of the preference attribute is "conflicting", the expert may change this value and affect the "semi-conflicting" and "consensual" values.

An ARFF file (the input file of Weka) is generated for each OLAP log file. It is used as an input for the used classification algorithms, namely, *ID3* which is a decision tree method based on the computation of entropy to generate the information gain and select attributes, *Classification Via Clustering* which is a simple meta-classifier that uses a cluster for classification. For cluster algorithms that use a fixed number of clusters, like SimpleKMeans, the user has to make sure that the number of clusters to generate are the same as the number of class labels in the dataset in order to obtain a useful model., *Multi class Classifier* which is a classification method handles multi-class datasets with 2-class distribution classifiers, *Hyperpipes* which is a classification algorithm constructed for each category; it contains all points of that category (essentially records the attribute bounds observed for each category); the test instances are classified according to the category that "most contains the instance", and *CVParameterSelection* which is a classification algorithm for performing parameter selection by cross-validation for any classifier.

Aiming to evaluate our clustering method, we apply the metrics usually of use:

- The **True Positive (TP)** rate is the proportion of examples which were classified as class x, among all examples which truly have class x, i.e. how much part of the class was captured. It is equivalent to **Recall**;
- The False Positive (FP) rate is the proportion of examples which were classified as class x, but belong to a different class, among all examples which are not of class x;
- The **Precision** is the proportion of the examples which truly have class x among all those which were classified as class x;
- The **F-Measure** is simply $(2*Precision*Recall)/(Precision+Recall)$ , a combined measure for precision and recall;
- The **Receiver operating characteristic** (ROC) is the relationship between the TP and FP rates.

Table 4: Results of classification with ten-fold cross validation

| Classification Method | TP Rate | FP Rate | Precision | Recall | F-Measure | ROC | Preference Class |
|---|---|---|---|---|---|---|---|
| ID3 | 1 | 0.004 | 0.996 | 1 | 0.998 | 0.996 | Conflicting |
| | 0.994 | 0 | 1 | 0.994 | 0.997 | 0.995 | Semi-Conflicting |
| | 0.996 | 0.001 | 0.998 | 0.996 | 0.997 | 0.997 | Consensual |
| Classification via clustering | 1 | 0.304 | 0.765 | 1 | 0.867 | 0.848 | Conflicting |
| | 0.595 | 0.1 | 0.666 | 0.595 | 0.628 | 0.747 | Semi-conflicting |
| | 0.398 | 0.034 | 0.797 | 0.398 | 0.531 | 0.682 | Consensual |
| Multiclass classifier | 1 | 1 | 0.498 | 1 | 0.665 | 0.5 | Conflicting |
| | 0 | 0 | 0 | 0 | 0 | 0.499 | Semi-Conflicting |
| | 0 | 0 | 0 | 0 | 0 | 0.499 | Consensual |
| Hyper pipes | 1 | 0.453 | 0.687 | 1 | 0.814 | 0.773 | Conflicting |
| | 0.994 | 0.033 | 1 | 0.909 | 0.994 | 0.997 | Semi-Conflicting |
| | 0 | 0 | 0 | 0 | 0 | 0.7 | Consensual |

Table 5: Results of classification with twenty-fold cross validation

| Classification Method | TP Rate | FP Rate | Precision | Recall | F-Measure | ROC | Preference Class |
|---|---|---|---|---|---|---|---|
| ID3 | 1 | 0.004 | 0.996 | 1 | 0.998 | 0.996 | Conflicting |
| | 0.994 | 0 | 1 | 0.994 | 0.997 | 0.995 | Semi-Conflicting |
| | 0.996 | 0.001 | 0.998 | 0.996 | 0.997 | 0.997 | consensual |
| Classification via clustering | 1 | 0.353 | 0.765 | 1 | 0.849 | 0.823 | Conflicting |
| | 0.349 | 0.017 | 0.666 | 0.349 | 0.499 | 0.666 | Semi-conflicting |
| | 0.648 | 0.083 | 0.797 | 0.648 | 0.683 | 0.782 | Consensual |
| Multiclass classifier | 1 | 1 | 0.498 | 1 | 0.665 | 0.498 | Conflicting |
| | 0 | 0 | 0 | 0 | 0 | 0.499 | Semi-Conflicting |
| | 0 | 0 | 0 | 0 | 0 | 0.497 | consensual |
| Hyper pipes | 1 | 0.478 | 0.675 | 1 | 0.806 | 0.761 | Conflicting |
| | 0.994 | 0.017 | 0.953 | 0.994 | 0.973 | 0.997 | Semi-Conflicting |
| | 0 | 0 | 0 | 0 | 0 | 0.683 | consensual |

In our experiments, as far as the value of the cross validation fold increases, the evaluation criteria produce better results and our preferences are correctly classified as illustrated by table 5. For evaluation of an error rate, we used the both of ten-fold cross validation and twenty-fold cross validation : all cases were randomly re-ordered, and then the set of all cases is divided into respectively ten and twenty mutually disjoint subsets of approximately equal size.

To assess the performance of our method, several classification methods were launched. As shown by the table 4, the ID3 algorithm engenders a precision equal to 99.6% for the first class, 100% for the second class and 99.8% for the third class. However, the classification via clustering technique generates a precision equal to 76.5 % for the first class, 66.6% for the second class and 79.7% for the third class. While MulticlassClassifier algorithm produces a precision equal to 49.8% only for the first class. Finally, Hyperpipes brings a precision equal to 68.5 % for the first class and 100% for the second class. Consequently, we stress out the accuracy of our proposed clustering method.

## 5. Conclusion

In this paper, we have proposed three stages to build and annotate analyst profile from OLAP log files starting, in a first stage, by the preprocessing of the log file which allows the text segmentation and the recognition of named entities. In a second stage, based on the conflict aspect, clustering of behavioral preferences in: (i) *consensual preferences*; (ii) *semi-conflicting preferences* and (iii) *conflicting preferences*. Then, enrichment of such behavioral preferences by adding personal and professional information. Finally, we may annotate the user profile-content using frequency.

There are different perspectives opened by this study. We think it would be interesting to confront the created profile and the spotted behavior. Moreover, we intend to investigate practical expressiveness of the derived user model. Finally, we plan to extend our contribution to personalize the query model.

# References

[1]. E. Ben Ahmed, A. Nabli, F. Gargouri, "A Survey of User-Centric Data Warehouses: From Personalization to Recommendation", The International Journal of Database Management Systems (IJDMS), May 2011, Volume 3, Number 2, 2011.

[2]. M. Bouzeghoub, D. Kostadinov, "Personnalisation de l'information: aperu de l'état de l'art et définition d'un modèle flexible de profils", In Proceedings of COnférence en Recherche d'Information et Applications (CORIA'05), pp. 201-218 , 2005.

[3]. V. Challam, S. Gauch, A. Chandramouli, "Contextual Search Using Ontology-Based User Profiles", In Proceedings of RIAO 2007, Pittsburgh USA, 2007.

[4]. M. Cherniack, E. Galvez, M. Franklin, S. Zdonik, "Profile-Driven Cache Management", In Proceedings of the 19th International Conference on Data Engineering, Bangalore, India , 2003.

[5]. O. Ferret, B. Grau, M. Hurault-Plantet, G. Illouz, C. Jacquemin, L. Monceaux, I. Robba, A. Vilnat, "How NLP Can Improve Question Answering", In Revue Knowledge Organization, 2002.

[6]. M. Golfarelli, S. Rizzi, "Expressing OLAP Preferences", Proceedings of the 21$^{st}$ International Conference on Scientific and Statistical Database Management, pp. 83-91, 2009.

[7]. H. Jerbi, F. Ravat, O. Teste, G. Zurfluh, "Management of context-aware preferences in multidimensional databases", International Conference on Digital Information Management (ICDIM'08),pp 669-675, 2008.

[8]. H. Jerbi, F. Ravat, O. Teste, G. Zurfluh, "Personnalisation du contenu des bases de données multidimensionnelles", Journées Francophones sur les Entrepôts de Données et l'Analyse en ligne (EDA'10), Djerba, Tunisie, pp. 520, 2010.

[9]. F. Liu, C. Yu, W. Meng, "Personalized Web Search For Improving Retrieval Effectiveness", IEEE Transactions on Knowledge and Data Engineering, vol. 16, n1, pp. 28-40, 2004.

[10]. J. P. Mc Gowan, "A multiple model approach to personalized information access, Master Thesis in Computer Science, Faculty of science, University College Dublin, 2003.

[11]. F. Ravat, O. Teste, "Personalization and OLAP Databases, Annals of Information Systems", New Trends in Data Warehousing and Data Analysis, Vol. 3, pp. 7192, 2008.

[12]. F. Ravat, O. Teste, G. Zurfluh, " Personnalisation de bases de données multidimensionnelles, INFORSID, pp. 121-136, 2007.

[13]. S. Rizzi, "OLAP preferences: a research agenda", International Workshop on Data Warehousing and OLAP (DOLAP07), pp. 99-100, 2007.

[14]. S. Rizzi, "New Frontiers in Business Intelligence: Distribution and Personalization", Advances in Databases and Information Systems (ADBIS'10), pp. 23-30, 2010.

[15]. G. Salton, "Automatic Text Processing", Addison-Wesley Publishing Company, 1988.

[16]. A. Sieg, B. Mobasher, R. Burke, G. Prabu, S. Lytinen, "Representing user information context with ontologies", uahci05, 2005.

[17]. A. Sieg, B. Mobasher, S. Lytinen, R. Burke, "Using Concept Hierarchies to Enhance User Queries in Web-based Information Retrieval", Artificial Intelligence and Applications (AIA), 2004.

**Eya Ben Ahmed** is carrying out a PhD degree in Computer Sciences in Sfax University in Tunisia. Her research is focused on data mining techniques and data warehouses.

**Ahlem Nabli** obtained her Ph.D. in Computer Science from Sfax University in Tunisia in 2010. She is currently an assistant professor of Computer Science at the Faculty of Sciences of Sfax in Tunisia. Her research interests include data warehouses and ontologies.

**Faïez Gargouri** obtained his Ph.D. in Renes Descartes, Paris 5 in 1995. He is actually a Professor of Computer Science at the Higher Institute of Computer Science and Multimedia of Sfax in Tunisia. His research interests include Information system, ontology engineering, semantic web, advanced databases, and data warehouses.